\documentclass[aps,pre,twocolumn,amsmath,amssymb,showpacs,amsfonts]{revtex4}
\usepackage{epsfig}
\usepackage{graphicx}
\usepackage{dcolumn}
\usepackage{bm}

\begin{document}

\title{
Evolution to a singular measure and two sums of Lyapunov exponents}
\author{Itzhak Fouxon}
\affiliation{Raymond and Beverly Sackler School of Physics and Astronomy,
Tel-Aviv University, Tel-Aviv 69978, Israel}
\begin{abstract}

We consider dissipative dynamical systems represented by a smooth compressible flow in a finite domain. The density
evolves according to the continuity (Liouville) equation. For a general, non-degenerate flow the result of the infinite
time evolution of an initially smooth density is a singular measure. We give a condition for the non-degeneracy which allows to
decide for a given flow whether the infinite time limit is singular. The condition uses a Green-Kubo type formula for the
space-averaged sum of forward and backward-in-time Lyapunov exponents.
We discuss how the sums determine the fluctuations of the entropy production rate in the SRB state and
give examples of computation of the sums for certain velocity fields.

\end{abstract}
 \maketitle

Dissipative dynamical systems are one of the main paradigms of non-equilibrium physics \cite{Dorfman}.
Within the paradigm the system is represented by a point in a $d-$dimensional "phase space". The velocity of this point is the local value of a
prescribed smooth velocity field. The latter is compressible so the phase-space volume and the Gibbs entropy are not conserved. This setting allows to describe effectively non-equilibrium situations with exchange of entropy with the environment. An example of such a system, that played an important role in the development of understanding of physics far from equilibrium, is the Lorenz oscillator \cite{Lorenz}. This system clarifies a rather
general feature of the dissipative systems - at large times the system's trajectories asymptote a zero-volume set in space (attractor). The stationary phase-space measure (sometimes given by "the SRB measure" \cite{Ruelle1,Dorfman,Sinai,Bowen}) $n_{st}$ is supported on the attractor and is singular: it is zero everywhere except on the attractor where it is infinite. Here we consider a general dissipative system and
study the formation of singularities, both infinities and zeros, as a result of evolution of an initially smooth density. We show that whether infinities or
zeros are formed is decided by the degeneracy of two time integrals of a pair-correlation function of velocity divergence. The divergence is sometimes referred to as the entropy production rate. These integrals correspond to the exact Green-Kubo type formulas for the sums of forward- and backward-in-time Lyapunov exponents respectively, where the second formula generalizes the previous result in \cite{FF}. The same quantities are shown to characterize the fluctuations of the entropy production rate in the steady state. A explicit formula for the sums' difference is provided. An essential feature of our treatment is the use of Lagrangian trajectories approach that allowed advancement in a whole range of problems recently, see \cite{GFV} and references therein.
The results apply to flows in a finite domain that either do not depend on time explicitly or depend on time but are stationary with respect to statistics defined by the spatial averaging (in the latter case $n_{st}$ evolves in time and is stationary only statistically).

Generally a Lyapunov exponent is a non-trivial functional of the velocity that does not allow for a treatable expression in terms of that velocity. As our Green-Kubo type formulas show, exceptionally, the sums of Lyapunov exponents can be expressed efficiently in terms of the velocity. The formulas allow a much better control over the sum than over any other combination of the Lyapunov exponents. Since the sums play a crucial role in determining the long-time behavior, then the ability to evaluate them is very important. The use of the Green-Kubo type formula for the sum of Lyapunov exponents was made in the context of studying the behavior of the pollutants in the wave turbulence, see \cite{FFV,FV} and also \cite{BFS}, and in analyzing the limit
of the weakly compressible dynamics \cite{IF2}. Here, as an example, the formulas are applied to a random velocity with short correlation time.

We consider a generic dynamical system governed by a deterministic velocity field $\bm v(\bm r)$ defined over some "phase space" with a $d-$dimensional coordinate $\bm r$. It will be assumed that the total volume of the available phase space is finite and it is
preserved by the flow (set equal to unity). The trajectories in the phase space are defined by
\begin{eqnarray}&&
\partial_t \bm q(t, \bm r)=\bm v[\bm q(t, \bm r)],\ \ \bm q(0, \bm r)=\bm r. \label{basic}
\end{eqnarray}
It is assumed that $\bm v(\bm r)$ is smooth so that the solution to the above system is well-defined and unique. We will
consider both $t>0$ and $t<0$ in the above equation.

We study dissipative systems for which $\omega (\bm r)\equiv \nabla \cdot\bm v(\bm r)$ is generally non-vanishing.
If the trajectories approach a zero-volume attractor at large times, then, clearly, under the dynamical evolution the volume of finite regions of the phase space decays to zero (this does not contradict the conservation of the total volume, see the third page).
In particular, the volume of initially infinitesimal regions approaches zero at large times. It is thus instructive to consider the evolution of
an infinitesimal volume $V(t)$ located initially near the point $\bm r$. This obeys
\cite{Batchelor},
\begin{eqnarray}&&
\!\frac{d\ln V}{dt}=\omega [\bm q(t, \bm r)],\ \ \ln \frac{V(t)}{V(0)}=\int_0^t \omega [\bm q(t', \bm r)]dt', \label{volume}
\end{eqnarray}
which holds as long as the linear dimensions of the volume are much smaller than the scale $\eta$ of smoothness of $\bm v$.
The ratio $V(t)/V(0)$ is the jacobian $J(t, \bm r)$ of $\bm q(t, \bm r)$,
\begin{eqnarray}&&
\!\!\!\!\!\!\frac{V(t)}{V(0)}=J(t, \bm r)\equiv \det \frac{\partial \bm q(t, \bm r)}{\partial \bm r}=\exp\left[\int_0^t \omega [\bm q(t', \bm r)]dt'\right].\nonumber
\end{eqnarray}
It determines the sum of the Lyapunov exponents $\lambda_i$ as 
\begin{eqnarray}&&
\sum \lambda_i(\bm r)=\lim_{t\to\infty}\frac{1}{t}\ln \frac{V(t)}{V(0)}= \lim_{t\to\infty}\frac{1}{t}
\int_0^t \omega [\bm q(t', \bm r)]dt'.\nonumber
\end{eqnarray}
Thus $\sum \lambda_i<0$ would make one expect that initial volumes decay exponentially
to zero at large times. This is however beyond this analysis: for any finite initial volume, however small,
one of the dimensions will eventually become larger than $\eta$ and Eq.~(\ref{volume}) will break down. Here we assume positivity
of the principal Lyapunov exponent $\lambda_1$ (for $\lambda_1<0$ the evolution transforms the phase space into a collection of point masses).

To address the global evolution of the phase space, we consider the phase space density $n(t, \bm r)$. Its large-time limit at a fixed point $\bm r$ is determined by the limiting value of the ratio of the volumes for evolution backward-in-time and not forward-in-time,
\begin{eqnarray}&&
\!\!\!\!\!\!\!\!\!\!\!\!\!\!\sum \lambda^-_i(\bm r)\!=\!\!\lim_{T\to\infty}\!\frac{1}{T}\ln\! \frac{V(-T)}{V(0)}
\!=\!-\!\!\lim_{T\to\infty}\!
\int_{-T}^0\!\!\! \!\frac{\omega [\bm q(t, \bm r)] dt}{T}, \label{sumdefinition}
\end{eqnarray}
where $\lambda_i^{-}(\bm r)$ are the Lyapunov exponents of the backward-in-time evolution $\bm q(-t, \bm r)$.
Density, like any transported field \cite{GFV}, is governed by the trajectories propagation backward (and not forward) in time.
To find $n(t, \bm r)$ one needs to find the trajectory that arrives at $\bm r$ at time $t$ or, equivalently, consider the evolution of the trajectory
backward-in-time starting from that point. The Lyapunov exponents for backward-in-time evolution were discussed in this context in
\cite{BFF}. We give a self-contained presentation avoiding some details not important here. The density obeys
the continuity (Liouville) equation
\begin{eqnarray}&&
\partial_t n+\nabla\cdot[n\bm v]=0, \label{continuity}
\end{eqnarray}
describing mass conservation $n[t, \bm q(t, \bm r)]J(t, \bm r)=n(0, \bm r)$. Evolution of density in a moving frame $\bm q(t, \bm r)$ obeys
\begin{eqnarray}&&
\lim_{t\to\infty}\frac{1}{t}\ln\frac{n[t, \bm q(t, \bm r)]}{n(0, \bm r)}=-\sum  \lambda_i(\bm r). \label{d10}
\end{eqnarray}
To discuss the evolution of density at ia fixed point in space we set up the initial condition in the remote past at $t=-T$ and
study $n(\bm r)\equiv n(0, \bm r)$ at $T\to\infty$,
\begin{eqnarray}&&
\!\!\!\!\!\!\!\!n(\bm r)=n[-T, \bm q(-T, \bm r)]\exp\left[-\int_{-T}^0 \omega [\bm q(t', \bm r)]dt'\right],\\&&
\lim_{T\to\infty}\frac{1}{T}\ln\frac{n(\bm r)}{n[-T, \bm q(-T, \bm r)]}=\sum  \lambda_i^{-}(\bm r). \label{d1}
\end{eqnarray}
Both the logarithmic growth rate of density in the moving frame and at a fixed point are determined by the time-averages of
$\omega$, however the former by the average with respect to the evolution forward-in-time and the latter by the
evolution backward-in-time. These evolutions are different and generally $\sum  \lambda_i^{-}\neq \sum  \lambda_i$, see
\cite{BFF} and below. The difference is made explicit with the help of the Green-Kubo type representations for the sums. We use
\begin{eqnarray}&&
[\bm v(\bm r)\cdot \nabla] \bm q(t, \bm r)=\bm v[\bm q(t, \bm r)],  \label{sasha}
\end{eqnarray}
as can be proved by time differentiation \cite{FF}. Note that if one considers trajectories that issue from the initial points
$\bm r$ and $\bm r+\bm v(\bm r)\Delta t$ then at time $t$ they will be located at $\bm q(t, \bm r)$ and $\bm q(t+\Delta t, \bm r)$ because the second trajectory is the trajectory issuing from $\bm r$ shifted by the time interval $\Delta t$. Thus the distance between these trajectories is on the one hand $\partial \bm q/\partial \bm r$ times the initial separation vector $\bm v(\bm r)\Delta t$ and on the other hand it is
$\bm q(t+\Delta t, \bm r)-\bm q(t, \bm r)\approx \bm v[\bm q(t, \bm r)] \Delta t$. Comparing the two expressions one gets Eq.~(\ref{sasha}).
Using the latter equation one finds
\begin{eqnarray}&&
\frac{\partial}{\partial t} \omega [\bm q(t, \bm r)]=v_i[\bm q(t, \bm r)]\frac{\partial \omega(\bm x)}{\partial x_i}|_{\bm x=\bm q(t, \bm r)}=
\frac{\partial q_i(t, \bm r)}{\partial r_j}\nonumber\\&&  \times v_j(\bm r)\frac{\partial \omega(\bm x)}{\partial x_i}|_{\bm x=\bm q(t, \bm r)}=\left[v_j(\bm r)
\frac{\partial}{\partial r_j} \right]\omega [\bm q(t, \bm r)].
\end{eqnarray}
We now average over some spatial volume $\Omega$,
\begin{eqnarray}&&
\frac{d}{dt} \int \frac{d\bm r}{\Omega}\omega [\bm q(t, \bm r)]=-\int \frac{d\bm r}{\Omega}\omega [\bm r]\omega [\bm q(t, \bm r)]\nonumber\\&& +
\int \omega [\bm q(t, \bm r)] \bm v(\bm r)\cdot d\bm S,
\end{eqnarray}
where the last term is the surface integral over the region boundary. Taking $\Omega$ to be the whole space
the integral over the boundary vanishes and one finds
\begin{eqnarray}&&
\frac{d}{dt} \int d\bm r\omega [\bm q(t, \bm r)]=-\langle \omega(0)\omega(t)\rangle, \label{c1}\\&&
\langle \omega(0)\omega(t)\rangle\equiv \int d\bm r \omega [\bm r]\omega [\bm q(t, \bm r)].
\label{definition}
\end{eqnarray}
The above relation is similar to the relations with transient-time correlation functions as described in \cite{EvansMorris} and references therein.
Integrating Eq.~(\ref{c1}) we find
\begin{eqnarray}&&
\int d\bm r \omega [\bm q(t, \bm r)]=-\int_0^t \langle \omega(0)\omega(t')\rangle dt',
\end{eqnarray}
where we used that the LHS vanishes at $t=0$ as an integral of the derivative. Assuming the integral on the RHS
converges in the limit $|t|\to\infty$ (remind that $t$ above can be both positive and negative) we conclude that,
\begin{eqnarray}&&
\!\!\!\!\!\lim_{t\to\pm\infty}\int d\bm r\omega [\bm q(t, \bm r)]=-\int_0^{\pm\infty} \langle \omega(0)\omega(t)\rangle dt. \label{R1}
\end{eqnarray}
Averaging 
over time, exchanging the order of integrals and designating spatial averages by angular brackets,
\begin{eqnarray}&&
\left\langle \sum \lambda_i \right\rangle =\!-\!\int_0^{\infty} \!\!\langle \omega(0)\omega(t)\rangle dt, \label{b1}\\&&
\left\langle \sum \lambda_i^- \right\rangle =\!-\!\int_{-\infty}^0\!\!\! \langle \omega(0)\omega(t)\rangle dt.\label{b2}
\end{eqnarray}
The above relation depends only on the assumption of convergence of the integrals on the RHS. If the conditions of the Oseledets theorem \cite{Oseledets}
are satisfied, then $\sum \lambda_i(\bm r)$ and $\sum \lambda_i^-(\bm r)$ are constants ($\mu$ and $\mu^-$ respectively) for almost every $\bm r$ and
Eq.~(\ref{b2}) implies
\begin{eqnarray}&&
\mu^-=-\int_{-\infty}^0 \langle  \omega(0)\omega(t)\rangle dt. \label{equation}
\end{eqnarray}
This is the counterpart of the relation for $\mu$,
\begin{eqnarray}&&
\mu =-\int_0^{\infty} \langle \omega(0)\omega(t)\rangle dt,\label{equation1}
\end{eqnarray}
derived in \cite{FF}. The result can be generalized from time-independent velocity $\bm v(\bm r)$ to time-dependent velocity $\bm v(t, \bm r)$
which is stationary with respect to the statistics defined by spatial averaging. One demands equalities like
\begin{eqnarray}&&
\int d\bm r\omega(t, \bm r)\exp\left[-\int_0^t\omega[t', \bm q(t'|t, \bm r)]dt'\right]=\nonumber\\&&
\int d\bm r \omega(0, \bm r)\exp\left[-\int_{-t}^0\omega[t', \bm q(t'| \bm r)]dt'\right].
\end{eqnarray}
where $\bm q(t'|t, \bm r)$ is the trajectory that passes through $\bm r$ at time $t$.
The proof follows the lines of \cite{FF} exactly and does not need to be repeated here.
The formulas (\ref{equation})-(\ref{equation1}) for time-independent velocities are a particular case of the Kawasaki representation \cite{Kawasaki,EvansMorris}.

Based on Eqs.~(\ref{b1})-(\ref{b2}) one expects $\langle \sum\lambda_i \rangle$ and $\langle \sum \lambda_i^- \rangle$ are non-positive. However this does not follow from the non-negativity of the spectrum as the averaging is not with respect to the invariant measure.
Also generally $\langle \omega(0)\omega(t)\rangle$ is not an even function of $t$ and $\sum \lambda_i\neq \sum \lambda_i^-$.
It was proved in \cite{Ruelle,BFF,FF} that $\langle\sum \lambda_i\rangle\leq 0$. Since
$\lambda_i^-$ are the Lyapunov exponents of the time-reversed velocity field $-\bm v(-t, \bm r)$, then
also $\langle\sum \lambda_i^-\rangle\leq 0$. Note $\langle\sum \lambda_i^-\rangle >0$ would mean $\sum \lambda_i^-(\bm r)>0$ for
a finite volume, implying by Eq.~(\ref{d1}) infinite mass in that region and contradicting mass conservation (thus
$\langle\sum \lambda_i^-\rangle =0$ implies $\sum \lambda_i^-(\bm r)=0$ almost everywhere and the same for $\sum \lambda_i$).

In the non-degenerate case $\langle\sum \lambda_i\rangle$ and $\langle\sum \lambda_i^-\rangle$ are negative. An important case with degeneracy is
where $\bm v$ allows for a smooth solution to the stationary continuity
equation $\nabla\cdot[n_{st}\bm v]=0$ for the invariant density $n_{st}$. Here
\begin{eqnarray}&&
\!\!\!\!\!\!\omega[\bm q(t, \bm r)]=-\bm v\nabla \ln n_{st}|_{\bm q(t, \bm r)}=-\frac{d}{dt} \ln n_{st}[\bm q(t, \bm r)].\nonumber
\end{eqnarray}
Thus $\omega[\bm q(t, \bm r)]$ is representable as a time-derivative of a bounded function. It follows that its time-averages $\sum \lambda_i(\bm r)$ and $\sum \lambda_i^-(\bm r)$ vanish (also $n_{st}$ is the same for backward- and forward-in-time dynamics \cite{Ruelle}).
In the generic case of $\langle\sum \lambda_i\rangle<0$  and $\langle\sum \lambda_i^-\rangle<0$ Eqs.~(\ref{equation})-(\ref{equation1})
give complementary descriptions of the formation of singularities in the limit of infinite evolution time. The inequality
$\langle\sum \lambda_i^-\rangle<0$ implies the density decays to zero in a finite region of space.
(Thus roughly $\langle\sum \lambda_i^-\rangle$ is the rate at which trajectories leave the regions outside the attractor.)
Complementarily, $\langle\sum \lambda_i\rangle<0$ implies the density becomes infinite for almost every trajectory starting in a finite region of space. (This corresponds to the accumulation of different trajectories on a zero-volume attractor). Nothing seems to forbid $\langle\sum \lambda_i\rangle=0$ at $\langle\sum \lambda_i^-\rangle<0$ or $\langle\sum \lambda_i^-\rangle<0$ at
$\langle\sum \lambda_i\rangle=0$ (say the rate of density vanishing can be non-exponential). Thus to decide whether a given compressible flow will produce singularities out of initially smooth density one has to check degeneracy of $\int_{-\infty}^0
\langle \omega(0)\omega(t)\rangle dt$ and $\int_0^{-\infty} \langle \omega(0)\omega(t)\rangle dt$.

We analyze the difference between $\langle\sum \lambda_i\rangle$ and $\langle\sum \lambda_i^-\rangle$. Using Eq.~(\ref{sumdefinition}) we express $\langle\sum \lambda_i^-\rangle$ via the forward-in-time evolution as
\begin{eqnarray}&&
\!\!\!\!\!\!\left\langle\sum \lambda_i^-\right\rangle=-\lim_{T\to\infty}
\int_{-T}^0 \frac{dt}{T} \int d\bm r \omega [\bm q(t, \bm r)]
=\!-\!\lim_{T\to\infty}\nonumber\\&&
\!\!\! \!\!\!\int_0^{T}\!\! \frac{dt}{T}  \int d\bm x \omega [\bm q(t, \bm x)]\exp\left[\int_0^T \omega [\bm q(t', \bm x)]dt'\right],\label{back}
\end{eqnarray}
where we changed variables $\bm x=\bm q(-T, \bm r)$ and shifted the time-variable. If $\omega [\bm q(t, \bm r)]$ has a finite correlation
$\tau_c$, as defined by spatial averaging, then the cumulants of
$\int_0^T \omega [\bm q(t', \bm r)]dt'$ are proportional to $T$ at $T\gg \tau_c$ and
\begin{eqnarray}&&
\gamma(b)\equiv \lim_{T\to\infty}(1/T) \ln \int d\bm r \exp\left[b\int_0^T \omega [\bm q(t', \bm r)]dt'\right],\nonumber
\end{eqnarray}
is a  non-trivial function, as follows from the cumulant expansion theorem \cite{Ma}. From  Eq.~(\ref{back}) and the conservation of the total volume $\gamma(1)=0$ we have
\begin{eqnarray}&&
\gamma'(0)=\left\langle\sum \lambda_i\right\rangle,\ \ \gamma'(1)=-\left\langle\sum \lambda_i^-\right\rangle.
\end{eqnarray}
Holder inequality implies $\gamma(b)$ is a convex function which together with $\gamma(0)=\gamma(1)=0$  implies
$\gamma'(0)\leq 0$ and $\gamma'(1)\geq 0$. This is an alternative proof $\langle\sum \lambda_i\rangle$ and $\langle\sum \lambda_i^-\rangle$ are non-positive. Further, at $t\gg \tau_c$ the probability density $P(x, t)$ of a "sum" $x(t)\equiv\int_0^t \omega [\bm q(t', \bm r)]dt'/t$ of a large number
$\sim t/\tau_c$ of independent random variables has a universal form $P(x, t)\sim \exp[-tS(x)]$. Here the "entropy" function $S(x)$ is the Legendre
transform of $\gamma(b)$. It is positive everywhere except for $x=\mu$, thus describing the exponential decay of the volume fraction of regions for which $x(t)$ deviates from its long-time limit $\mu$ (here constant $\sum\lambda_i(\bm r)$ is assumed). The volume conservation relation $\int \exp(t[x-S(x)])\sim 1$
is determined by $x_*$ giving minimum to $x-S(x)$ and obeying $x_*=S(x_*)$ due to $\gamma(1)=0$. While at $t\to\infty$ the volume of the density support generally decreases, at any finite $t$ it equals one. Evolution of a fraction of space $\exp[-tS(x_*)]$ expands it by the factor $\exp[tx_*]$ and its volume roughly equals one.

Often, like for the Galavotti-Cohen relation \cite{GallavottiCohen}, one identifies
$\sigma \equiv -\int_0^T \omega [\bm q(t, \bm x)]dt/T$ as the entropy production rate in the non-equilibrium steady state described by $n_{st}$ (if it exists).
Fluctuations of $\sigma$ in the steady state are described by the large deviations function obtainable as the Legendre
transform of ${\tilde \gamma}(b)$,
\begin{eqnarray}&&
\!\!\!\!\!\!\!\!{\tilde \gamma}(b)\equiv \lim_{T\to\infty}\frac{1}{T} \ln \int n_{st}(\bm r)d\bm r \exp\left[b\int_0^T \omega [\bm q(t', \bm r)]dt'\right].\nonumber
\end{eqnarray}
Let us show that the equality between $\gamma(b)$ and ${\tilde \gamma}(b)$ holds quite generally, cf. \cite{FF,Gawedzki}. We introduce a time $t_*$ obeying $0<t_*\ll T$ and make a change of variables $\bm x=\bm q(t_*, \bm r)$ in the definition of $\gamma(b)$. We find
\begin{eqnarray}&&
\!\!\!\!\!\!\!\!\gamma(b)\!=\!\lim_{T\to\infty}\frac{1}{T} \ln \int n(t_*, \bm x) d\bm x \exp\Biggl[b\int_{-t_*}^{T-t_*} \omega [\bm q(t', \bm x)]dt' \Biggr]
\nonumber\\&&\!\!\!\!\!\!\!\!\approx \lim_{T\to\infty}\frac{1}{T} \ln \int n(t_*, \bm x) d\bm x \exp\left[b\int_0^{T} \omega [\bm q(t', \bm x)]dt' \right]
,\nonumber
\end{eqnarray}
where $n(t, \bm x)$ obeys Eq.~(\ref{continuity}) with $n(0, \bm x)=1 $. We shifted the time variables in the integral and used $t_*\ll T$. Taking now the limits $t_*\to\infty$ and $T\to \infty$
at $t_*\ll T$ we obtain $\gamma(b)={\tilde \gamma}(b)$. The main assumption is that the density $n(t_*, \bm x)$ tends to the stationary measure
at large $t_*$, cf. \cite{IF2}.
Thus the relations derived previously for $\gamma(b)$ also characterize the fluctuations of the entropy production rate in the
steady state. All the above relations can be generalized to random, time-dependent velocity \cite{BFF,FF}.

The difference between $\langle\sum \lambda_i\rangle$ and $\langle\sum \lambda_i^-\rangle$ is due to non-even part of $\langle \omega(0)\omega(t)\rangle$. This vanishes
for time-reversible statistics.
One can derive
a series representation for the difference in terms of different time correlations of $\omega$ on the trajectory $\bm q(t, \bm r)$.
Change of variables $\bm x=\bm q(-t, \bm r)$ in the definition (\ref{definition}) of $\langle \omega(0)\omega(t)\rangle$ gives
\begin{eqnarray}&&
\!\!\!\!\!\!\!\!\langle \omega(0)\omega(-t)\rangle\!=\!\int d\bm r\omega(\bm r)\omega [\bm q(t, \bm r)]\exp\left[\int_0^{t} \omega [\bm q(t', \bm r)]dt' \right].\nonumber
\end{eqnarray}
Expanding the exponent into series we find ($t_0\equiv t$)
\begin{eqnarray}&&
\!\!\!\!\!\!\!\!\left\langle\left[\sum \lambda_i-\sum \lambda_i^-\right]\right\rangle\!=
\sum_{n=1}^{\infty}
\int_0^{\infty}dt\int_0^t dt_1..\int_0^{t_{n-1}} dt_n
\nonumber\\&& \int d\bm r\omega (\bm r)\omega [\bm q(t_n, \bm r)]..\omega [\bm q(t_1, \bm r)]
\omega [\bm q(t, \bm r)].\nonumber
\end{eqnarray}
Eqs.~(\ref{equation})-(\ref{equation1}) allow to compute space-averaged sums of the Lyapunov exponents easier than the Lyapunov exponents themselves. The latter involve correlation functions of all orders, not just the
second one. For example consider a random velocity field short-correlated in time.
Then $\int \langle \omega(0)\omega(t)\rangle dt$ is determined by times where $\bm q(t, \bm r)\approx \bm r$,
\begin{eqnarray}&&
\left\langle\sum \lambda_i\right\rangle=\left\langle\sum \lambda_i^-\right\rangle\!=\!-(1/2)\int \langle \omega(0, \bm r)\omega(t, \bm r)\rangle dt, \nonumber
\end{eqnarray}
where we used that the integrand is an even function of $t$.
For the so-called Kraichnan model $\bm v$ is a Gaussian velocity field with zero mean and pair-correlation $\langle
v_{\alpha}(t_1, \bm r_1)v_{\beta}(t_2, \bm r_2)\rangle =[V_0\delta_{\alpha\beta}-K_{\alpha\beta}(\bm r_2-\bm r_1)]\delta(t_2-t_1)$.
We find $\mu=\mu^-\!=\!(-1/2)\nabla_{\alpha}\nabla_{\beta}K_{\alpha\beta}(0)$, see e. g. \cite{BFF}. The formulas for other combinations of the Lyapunov exponents are more cumbersome \cite{BFF}. For isotropic ensemble $K_{\alpha\beta}(\bm r)=D[(d+1-2\Gamma)
\delta_{\alpha\beta}r^2+2(d\Gamma-1)r_{\alpha}r_{\beta}]$, where $D$ is a constant and $0\leq \Gamma\leq 1$ is a measure of the flow compressibility,
one obtains
$\mu=\mu^-=-dD\Gamma(d-1)(d+2)$.
Another use of  Eqs.~(\ref{equation})-(\ref{equation1}) is for small-amplitude waves such as surface waves in the ocean. Here higher powers of $\bm v$ are relatively small.  The calculation in \cite{BFS,FFV,FV} showed that up to the fourth order in the wave amplitude $\mu=0$. This is while $\lambda_1$ is of the fourth order in the wave amplitude. Thus, though the surface flow is a generic flow with compressibility of order one, there is a degeneracy that cancels the long-time effects of compressibility,
$\sum \lambda_i/\lambda_1\ll 1$. Thus how to explain clustering of pollutants on oceans' surfaces is unclear. Now, the difference between $\mu^-$ and $\mu$ raises the possibility
that $\mu^-$ does not vanish in the fourth order in the wave amplitude. The study of this possibility is a subject for future work.

I thank K. Gawedzki and M. Cencini for discussions. This work was supported by COST Action MP$0806$.



\end{document}